\documentclass[twocolumn,showpacs,preprintnumbers,amsmath,amssymb,longbibliography]{revtex4-1}

\usepackage{graphicx}
\usepackage{dcolumn}
\usepackage{bm}
\usepackage{liecolon}
\usepackage{nicefrac}
\usepackage{units}

\usepackage{multirow}

\begin{document}

\preprint{}

\title{Chromatic and Dispersive Effects in Nonlinear Integrable Optics}

\author{Stephen D. Webb}%
 \email{swebb@radiasoft.net}
 \homepage{www.radiasoft.net}
 \author{David L. Bruhwiler}
\affiliation{RadiaSoft, LLC, Boulder, CO 80304, USA}
\author{Alexander Valishev}
\author{Sergei Nagaitsev}
\affiliation{Fermi National Accelerator Laboratory, Batavia, Illinois 60510, USA}
\author{Viatcheslav V. Danilov}
\affiliation{Spallation Neutron Source, Oak Ridge National Laboratory, Oak Ridge, Tennessee 37830, USA}

\date{\today}

\begin{abstract}
Proton accumulator rings and other circular hadron accelerators are susceptible to intensity-driven parametric instabilities because the zero-current charged particle dynamics are characterized by a single tune. Landau damping can suppress these instabilities, which requires energy spread in the beam or introducing nonlinear magnets such as octupoles. However, this approach reduces dynamic aperture. Nonlinear integrable optics can suppress parametric instabilities independent of energy spread in the distribution, while preserving the dynamic aperture. This novel approach promises to reduce particle losses and enable order-of-magnitude increases in beam intensity. In this paper we present results, obtained using the Lie operator formalism, on how chromaticity and dispersion affect particle orbits in integrable optics. We conclude that chromaticity in general breaks the integrability, unless the vertical and horizontal chromaticities are equal. Because of this, the chromaticity correcting magnets can be weaker and fewer correcting magnet families are required, thus minimizing the impact on dynamic aperture.
\end{abstract}

\maketitle

\section{Introduction}

Modern accelerator applications require increasingly high intensity beams. For example, the European Spallation Source plans for a proton beam with \unit[5]{MW} average power. The Proton Improvement Plan at Fermilab, intended to drive neutrino experiments, will top \unit[1]{MW} with the ability to expand beyond that. As the beam power increases, the stability of the beam becomes threatened by coherent collective effects due to direct space charge -- the canonical example of this is beam halo. Mitigating these effects to reduce beam loss requires improvements in beam transport.

One proposed tool to mitigate coherent effects is the nonlinear integrable lattices~\cite{danilovNagaitsev:2010, nagValDan:2010} to introduce large transverse tune spreads while still maintaining bounded, regular orbits. The principle here is to construct an accelerator lattice which leads to bounded, regular motion in the transverse plane for on-momentum particles while having extremely broad tune spreads. The large tune spread with amplitude decoheres any oscillations which would normally drive coherent space charge instabilities~\cite{webb:12}.

These lattices have been studied in the two-dimensional regime for both single-particle and collective effects. Preliminary work indicates that they are robust against space charge and small perturbations in the transverse lattice design. However, some numerical studies have indicated susceptibility to off-momentum effects such as dispersion in the elliptic magnet sections and chromaticity in the lattice.

In this paper, we address chromaticity and dispersion effects using the Lie operator formalism. We obtain general results on how these effects break the integrability of the lattice, as well as how we may mitigate those effects. We conclude from our studies that having equal vertical and horizontal chromaticities and having no dispersion inside the drifts for the nonlinear elements restores the integrability exactly for a coasting beam. We also find that the standard chromaticity correction schemes will also work for these lattices. Because we need only make the chromaticities equal, rather than make them vanish or close to vanishing to control tune spread, we are free to select a family of correcting sextupoles, octupoles, \emph{etc.} which has the least effect on the dynamic aperture.

\section{Single-turn Map \& Normal Form Analysis}

To provide a sufficiently general treatment of the nonlinear integrable lattices, we turn to the Lie operator formalism of Dragt, Forest, and others~\cite{dragtFinn:1976, dragtForest:1983, dragt_text:2011}. This formalism is useful for obtaining invariants of the lattice, such as the Courant-Snyder invariants and their nonlinear generalizations. This formalism begins with the Poisson brackets for the equations of motion for a dynamical variable $z$:
\begin{equation}\label{hamiltoneqns}
\dot{z} = -\{H, z\}
\end{equation}
where $H$ is the Hamiltonion and the overdot indicates the derivative with respect to the independent variable. We may interpret the Hamiltonian as a Lie operator, $\lieop{H}$, which acts on $z$ as such:
\begin{equation}
\dot{z} = -\lieop{H} z
\end{equation}
The action of the Lie operator on a dynamical variable is to take the Poisson brackets of $H$ with that variable, so that $\lieop{H} * = \{H, *\}$. Under these conditions, the solution to the differential equation (\ref{hamiltoneqns}) is
\begin{equation}
z(t) = e^{- \lieop{H} s} z(0)
\end{equation}
assuming that $H$ is independent of the independent variable.

This approach offers a number of advantages for accelerator applications. The first is that you can multiply these operators together, whereas you cannot multiply Hamiltonians directly. This is very fruitful, as a storage ring can be thought of as a series of piecewise constant Hamiltonians for drifts, quadrupoles, etc. The resulting map is simply the product of all the individual maps, in order. Second, and particularly useful for this application, the Lie operator formalism provides a clear path for canonical transformations, such as the transformation to the Courant-Snyder parameterization. Finally, there is a perturbation series for computing invariants, courtesy of the Baker-Campbell-Hausdorff series. This perturbation series provides increasingly high-order invariants of the motion which inform the dynamic aperture.

One identity we will require is the similarity transformation identity. This allows us to move accelerator elements around, manipulating their order without approximation. The following identity is exact:
\begin{equation}
e^{\lieop{f}} e^{\lieop{g}} e^{-\lieop{f}} = \exp\left ( \lieop{e^{\lieop{f}}g} \right )
\end{equation}
What this allows us to do is take an operator and move it around another operator -- this is convenient for computing the linear transfer map for a ring with nonlinear elements in preparation for a perturbation theory calculation. Specifically, we consider
\begin{equation}
e^{-\lieop{h_1}} e^{-\lieop{h_2}}
\end{equation}
To move the $h_2$ operator past $h_1$, we simply insert the identity to its left, so that
\begin{equation} \label{similarityGame}
\begin{split}
e^{-\lieop{h_1}} e^{-\lieop{h_2}} &= e^{-\lieop{h_2}}e^{\lieop{h_2}} e^{-\lieop{h_1}} e^{-\lieop{h_2}} \\&= e^{-\lieop{h_2}} \exp  \left (-\lieop{ e^{-\lieop{h_1}} h_2} \right )
\end{split}
\end{equation}
We will use this technique extensively in the next section. Let us now consider how all this applies to a storage ring.

If we have a series of accelerator elements, with Hamiltonians denoted by $\{H_1, H_2, \dots, H_N\}$, then we may think of each $H_i$ as generating the motion of the particles through element $i$ of length $\ell_i$. The combination of all these maps yields $\mathcal{H}$, the Hamiltonian which generates the single-turn map. Thus, we define
\begin{equation}\label{map}
\mathcal{M} = e^{- \lieop{\mathcal{H}}C} = \prod_{i = 1}^N e^{-\lieop{H_i}\ell_i}
\end{equation}
This product gives the single turn map from element $1$ through the $i$ elements and back to element $1$. Thus, given a phase space co\"{o}rdinate $z_0$, after a single turn through this ring it will find itself at $z_1$. $\mathcal{M}$ is the \emph{single-turn map} for this ring. It should be clear that $\mathcal{M}$ contains all the single-particle dynamical information for the ring\footnote{There are general issues about whether $\mathcal{H}$ exists, or exists only asymptotically, but throughout this paper we assume that it does exist and does not have any singularities near resonances, for example. Numerical simulations indicate that the Hamiltonian for the nonlinear elliptic potential characterizes the ideal system as both it and the second invariant are conserved to a very high degree. Thus, for this problem the assumption of $\mathcal{H}$ existing seems valid.}.

Philosophically, the goal of the transfer map approach is to obtain invariants. For purely linear lattices, this yields the Courant-Snyder invariants; Lie perturbation theory treatments (see, for example, Forest~\cite{forest_text:1998} or Chao~\cite{chao_lieAlgebraNotes:2009}) yield their approximate extensions with nonlinearities. These are frequently asymptotic, but contain a great deal of useful information regardless. In particular, if a lattice has an invariant, then we can rewrite the transfer map as a co\"{o}rdinate transformation which contains all the information specific to the location in the ring times a pure rotation operator which is a property of the ring. The linear version of this is the Twiss parameterization, and then rotations in the normalized $x-p$ space.

This is the \emph{normal form} analysis. Suppose $\mathcal{A}$ is a Lie operator which represents a co\"{o}rdinate transformation, so that $\mathcal{A} z = \overline{z}$. Then the transfer map in eqn.~(\ref{map}) can be rewritten in the form
\begin{equation}
e^{-\lieop{\mathcal{H}}C} = \mathcal{A}^{-1} e^{-\lieop{\overline{H}} C} \mathcal{A}
\end{equation}
where $\mathcal{A}$ represents the position-dependent co\"{o}rdinate transformation to bring $z \mapsto \overline{z}$, and $\overline{H}$ has no explicit dependence on where in the ring we are.

As an explicit example, consider the single turn map for a purely linear ring in one transverse dimension. This is a series of thin(or not) quadrupoles and drifts, which after Lie concatenation yields the transfer map
\begin{equation}
\mathcal{M} = e^{-\frac{\mu}{2}\lieop{\gamma x^2 + 2 \alpha x p + \beta p^2}}
\end{equation}
It is straightforward to show that any Lie concatenation of quadratic elements takes this form. Now, to rewrite the single-turn Hamiltonian $\mathcal{H}$ in the ``simplest possible way". Intuitively, it would be nice to turn the tilted ellipse into a circle, which can be accomplished with the normalizing map
\begin{equation}
\mathcal{A}^{-1} = \left (\begin{array}{cc}
\nicefrac{1}{\sqrt{\beta}} & 0 \\
\nicefrac{\alpha}{\sqrt{\beta}} & \sqrt{\beta}
\end{array} \right )
\end{equation}
The normalizing map contains the Twiss parameters! Under this transformation, the normalized Hamiltonian is given simply by
\begin{equation}
\overline{H} = \frac{\mu}{2} \left ( \overline{p}^2 + \overline{x}^2 \right )
\end{equation}
We can obtain the tune(s) explicitly from looking at the rotation of the normal form co\"{o}rdinates under the action of the normalized Hamiltonian. More importantly, we have a global invariant, $\varepsilon = \overline{p}^2 + \overline{x}^2$, which tells us a great deal about the beam dynamics. In fact, $\varepsilon$ is the emittance. Because all of the ring dynamics are contained in the $e^{-\lieop{\overline{H}} C}$ operator, any function of $\overline{H}$ is conserved -- this is the concept of beam matching. For nonlinearities, $\overline{H}$ may be obtained either using the resonance basis or the Baker-Campbell-Hausdorff formula (see, for example, \S 9.4 of Chao's notes~\cite{chao_lieAlgebraNotes:2009} for a thorough discussion of the topic) to include higher order terms.

\section{A Hierarchy of Lattices}

For nonlinear integrable optics lattices, there is in a sense a three-tiered hierarchy of the lattice. It is important to discuss this hierarchy here to avoid confusion and gain a sound conceptual understanding of what is to come. The three tiers are: (1) the bare lattice; (2) the corrected lattice; and (3) the nonlinear integrable lattice.

Tier (1), the bare lattice, contains only the drifts, dipoles, and quadrupoles. This lattice includes a drift section with equal vertical and horizontal beta functions, where the nonlinear elliptic magnets will be located. This lattice is linear for on-momentum particles. The bare lattice includes also the vertical and horizontal chromaticity for off-momentum effects. We do not consider here the effects of synchrotron oscillations -- the beam is assumed coasting. The bare lattice dictates the Twiss parameters as well as the dispersion functions -- the normalizing map $\mathcal{A}^{-1}$ of the previous section.

Tier (2), the corrected lattice, contains all of the elements in the bare lattice, as well as any families of nonlinear elements used to correct the bare lattice. Elements that can appear in the corrected lattice include chromaticity-correcting sextupoles, octupoles, \emph{etc.} It is noted by Chao~\cite{chao_lieAlgebraNotes:2009}, for example, that these nonlinearities can modify the normalizing maps. However, we can nevertheless choose our co\"{o}rdinate transformations so that the normalizing map for the bare lattice remains unchanged.

The nonlinear integrable lattice, Tier (3), contains all of Tiers (1) and (2), as well as the nonlinear elliptic magnets generating the elliptic potential described in \S V of~\cite{danilovNagaitsev:2010}. These magnets are positioned in a drift with equal vertical and horizontal beta functions. In this paper, we consider the magnets to be a single, smoothly varying unit. In practice, the drifts containing the elliptic potential magnets will be broken into segments containing two drifts symmetrically sandwiching a constant field strength magnet. Preliminary work indicates that this configuration does not change the ideal case substantially -- with ideal alignment we find well less than 1\% variation in the two invariants described in~\cite{danilovNagaitsev:2010}.

It is important to note that these three Tiers do comprise a hierarchy. Thus, Tier (2) is affected by Tier (1), but not Tier (3). Thus, when we refer to chromaticity, we refer to chromaticity as a Tier (2) effect. The nonlinear elliptic element exists on top of that lattice, but does not contribute to this chromaticity under our definition. Similarly, the Tier (2) and Tier (3) lattices do not affect the Twiss parameters -- the choice of co\"{o}rdinate transformation is made to normalize the Tier (1) transfer map and is unaffected by the Tier (2) and Tier (3) lattices.

With this nomenclature established, we continue to discuss the four-fold symmetric design for a nonlinear integrable lattice, currently under development at Fermilab.

\section{A Lie Operator Treatment of the IOTA Lattice}

\begin{figure}
\includegraphics[scale=0.45]{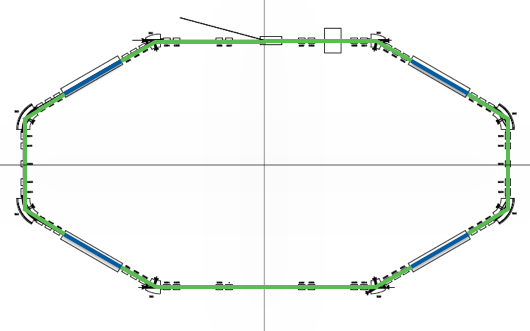}
\caption{The IOTA ring with nonlinear drift elements (blue) and the rest of the lattice (green), with a four-fold symmetry}\label{iotaring}
\end{figure}

We now have most of the key ideas in place to discuss the proposed elliptic lattice in the Integrable Optics Test Accelerator (IOTA) ring (figure~\ref{iotaring})~\cite{iota}. We will initially consider a coasting beam with no RF acceleration, so that the longitudinal momentum $\delta$ is a constant, and $z$ changes by $\eta_c \delta$ every turn. The longitudinal dynamics will not be of much interest in this section, except for the presence of energy spread for chromaticity.

The theory that follows applies to a single superperiod of an integrable nonlinear ring. In practice, a nonlinear integrable lattice  may include several superperiods. However, because of the periodicity, the analysis that follows still applies.

To represent this ring in the form of eqn.~\ref{map}, we need to contend with the strongly $s$-varying magnets. To do this, we imagine $N$ slices of constant magnetic field with length $\Delta s$, so that $N \Delta s = \ell$, the length of the drift. We can then take the limit of $N \rightarrow \infty$ with $N \Delta s$ fixed after our manipulations.

We consider the transfer map from the middle of an elliptic magnet segment to the middle of the next. Note that Lie operators are perhaps a little backwards from matrices. A string of matrices $N_1, N_2 \dots$ will be arrayed $\dots N_2 N_1 a$ when acting on a vector $a$. The matrices are resolved from right to left. Lie operators, which we will denote with script letters, resolve from left to right. Thus we would write $\mathcal{N}_1 \mathcal{N}_2 \dots \circ a$ to denote a string of operators which act on $a$ beginning with $\mathcal{N}_1$. Though contrary to what one might expect, this is the convention used and the one we follow in this paper.

\begin{figure}
\includegraphics[scale=0.75]{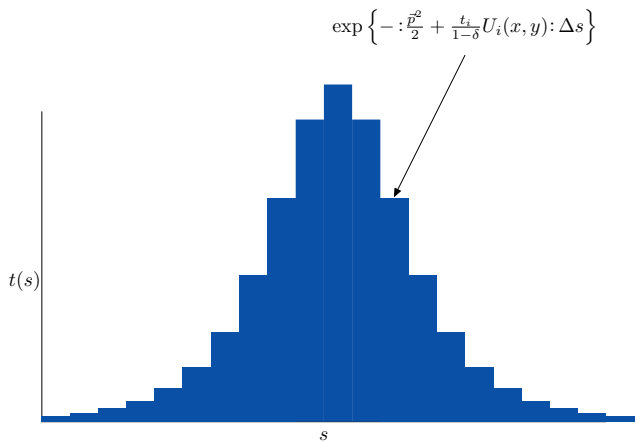}
\caption{Schematic of the piecewise constant elements for the elliptic magnet section. As $N \rightarrow \infty$ this becomes the ideal elliptic element.}\label{varyingStrength}
\end{figure}

It is convenient for exploiting the symmetry of the lattice to start in the middle of the elliptic magnet. Thus, starting with the middle of the elliptic magnet and working our way to the middle of the next, we have
\begin{equation}
\mathcal{M}_{0\rightarrow1} = \prod_{i = \nicefrac{N}{2} + 1}^N \exp \left \{ - \lieop{\frac{\vec{p}^2}{2} + \frac{t_i}{1 + \delta} U_i(x, y)} \Delta s \right \}
\end{equation}
This map imagines breaking up the second half of the elliptic magnet into $\nicefrac{N}{2}$ thin segments, each with a constant field strength. Each term in the above product moves a particle through the thin slice of constant field strength that it represents. For the ideal case of a smoothly varying elliptic magnet, we will take $N \rightarrow \infty$ and $\Delta s \rightarrow 0$, holding their product fixed. This factorization is shown schematically in figure~\ref{varyingStrength}.

This infinite product of maps takes us from the middle of the elliptic element to the edge. The map that brings us to the beginning of the next elliptic element we denote as
\begin{equation}
\mathcal{M}_{1\rightarrow2} = e^{-\lieop{h}}
\end{equation}
where $\lieop{h}$ contains all the quadrupoles, drifts, sextupoles for chromaticity correction, \emph{etc.} that describes the elements between the elliptic magnets. This is the Tier 2, corrected lattice surrounding the drift space surrounding the elliptic magnets and brings us from the end of the elliptic magnet to the beginning of the next. Because we began in the middle of a mirror-symmetric lattice, the second half of the elliptic magnet transfer map is just the mirror symmetric version of the first half. Thus, we have the transfer map explicitly as
\begin{equation}
\mathcal{M}_{2\rightarrow0} = \prod_{i = 0}^{\nicefrac{N}{2}} \exp \left \{ - \lieop{\frac{\vec{p}^2}{2} + \frac{t_i}{1 + \delta} U_i(x, y)} \Delta s \right \}
\end{equation}
The transfer map for the nonlinear integrable lattice is just the product of these individual maps, viz.
\begin{equation}
\mathcal{M}_{\textrm{IOTA}} = \mathcal{M}_{0\rightarrow1}\mathcal{M}_{1\rightarrow2}\mathcal{M}_{2\rightarrow0}
\end{equation}
The same argument applies for an IOTA lattice with a single symmetry, so long as our observing point starts in the middle of the elliptic magnet. We now want to find the normalizing co\"{o}rdinates for this lattice, as discussed in the previous section. We will look at this for the $\mathcal{M}_{2\rightarrow0}$ map -- the argument is symmetrical to the other half of the elliptic magnet.

We begin with a factoring that is second order in $\Delta s$ of each of the infinitesimal slices:
\begin{widetext}
\begin{equation}
\exp \left \{ - \lieop{\frac{\vec{p}^2}{2} + \frac{t_i}{1 + \delta} U_i(x, y)} \Delta s \right \} = e^{-\lieop{\frac{\vec{p}^2}{2}} \Delta s/2}\exp \left \{ - \lieop{\frac{t_i}{1 + \delta} U_i(x, y)} \Delta s \right \} e^{-\lieop{\frac{\vec{p}^2}{2}} \Delta s/2} + \mathcal{O}(\Delta s^3)
\end{equation}
\end{widetext}
Taken in the limit of $\Delta s \rightarrow 0$, this becomes exact. We now use the similarity transformation described by eqn.(\ref{similarityGame}), inserting the identity as drift operators to move the individual drifts to the left, resulting in the map taking the form
\begin{equation}
e^{-\lieop{\frac{\vec{p}^2}{2}} N \Delta s/2} \prod_{i = 0}^{\nicefrac{N}{2}} \exp \left \{ - \lieop{ \frac{t_i}{1 + \delta} e^{-\lieop{\vec{p}^2/2} (i + \nicefrac{1}{2}) \Delta s} U_i(x, y)} \right \}
\end{equation}
We have thus moved all of the linear parts of the elliptic magnet transfer map -- specifically the drift maps -- next to the rest of the lattice. This, along with its symmetrical argument for $\mathcal{M}_{0\rightarrow1}$, gives the transfer map for the nonlinear integrable IOTA lattice:
\begin{widetext}
\begin{equation} \label{iotaMap1}
\begin{split}
\mathcal{M}_{\textrm{IOTA}} = \prod_{i = \nicefrac{N}{2} + 1}^N \exp \left \{ - \lieop{ \frac{t_i}{1 + \delta} e^{-\lieop{\nicefrac{\vec{p}^2}{2}} (i - \nicefrac{N}{2} + \nicefrac{1}{2})\Delta s} U_i(x, y) \Delta s}  \right \} \times\\
\left ( e^{-\lieop{\frac{\vec{p}^2}{2}} \nicefrac{\ell}{2} }e^{-\lieop{h}} e^{-\lieop{\frac{\vec{p}^2}{2}} \nicefrac{\ell}{2}}\right )& \times\\
\prod_{i = 0}^{\nicefrac{N}{2}} \exp &\left \{ - \lieop{ \frac{t_i}{1 + \delta} e^{-\lieop{\nicefrac{\vec{p}^2}{2}} (i + \nicefrac{1}{2}) \Delta s} U_i(x, y) \Delta s} \right \}
\end{split}
\end{equation}
\end{widetext}
This factors the nonlinear integrable lattice into the corrected IOTA lattice transfer map, and the strongly nonlinear elliptic potential.  We analyze the corrected IOTA lattice first, as this is where we will obtain the normalizing maps $\mathcal{A}$, the bare lattice tunes, and the chromaticity.

\section{The Corrected IOTA Lattice}

In the previous section, we factored the transfer map into the nonlinear elliptic potential and the corrected lattice. The bare lattice with all nonlinear terms neglected contains all the information for computing the normalizing map. As discussed, the normalizing map contains the Twiss parameters, specifically the beta functions which carefully cancel to produce the integrable elliptic Hamiltonian. To describe this cancellation in a more general form to obtain effects due to dispersion and chromaticity, we need to carry out a normal form treatment of the IOTA lattice. We therefore consider the corrected IOTA lattice transfer map
\begin{equation}
\mathcal{M}_0 = e^{-\lieop{\frac{p^2}{2}} \nicefrac{\ell}{2}} e^{-\lieop{h}}e^{-\lieop{\frac{p^2}{2}} \nicefrac{\ell}{2} }
\end{equation}
This is the full Tier 2 lattice for the IOTA ring, in the absence of any elliptic elements. Thus, it contains the combined transfer maps for quadrupoles, drifts, and bends, as well as any chromaticity correcting elements such as sextupole families. The familiarity with chromaticity correction schemes means we push the details of this discussion to the appendix and instead approach the problem as computing the single-turn invariant Hamiltonian in the normalized co\"{o}rdinates.

Thus, we start from the assumption that the total Hamiltonian $H$ has been computed by some manipulation of Lie operators representing the individual elements, viz.
\begin{equation}\label{iotaProd}
\mathcal{M}_0 = e^{-\lieop{H_0} C} = \prod_{i = 0}^M e^{-\lieop{h_i} \ell_i}
\end{equation}
and furthermore that we have computed the normalized variables so that we may represent this as
\begin{equation} \label{correctedLattice}
e^{-\lieop{H_0}C} = \mathcal{A}^{-1} e^{-\lieop{\overline{H}}} \mathcal{A}
\end{equation}
Recall the peculiar choice of ordering for Lie operators, which act left-to-right instead of right-to-left. Thus, if $\mathcal{M}_0 \circ z$, then we would have the $h_0$ operator act on $z$ first, then the $h_1$, and on. We assume that $\mathcal{A}$ is derived purely from the linear part of the bare lattice -- the perturbation theory used to compute the full Hamiltonian $\overline{H}$ generates higher order, nonlinear normalizing maps if there are nonlinear elements in the corrected lattice. However, these can be swept into the form in eqn.~(\ref{correctedLattice}) with a judicious use of the similarity transformation technique described in eqn.~(\ref{similarityGame}).

The typical Courant-Snyder parameterization is a linear transformation on the $p$'s and $q$'s given by the matrix
\begin{equation}
\mathcal{A}^{-1} = \left (
\begin{array}{cccccc}
\nicefrac{1}{\sqrt{\beta_x}} & 0 & 0 & 0 & 0 & - \nicefrac{\eta}{\sqrt{\beta_x}} \\
\nicefrac{\alpha_x}{\sqrt{\beta_x}} & \sqrt{\beta_x} & 0 & 0 & 0 & - \nicefrac{\alpha_x \eta + \beta_x \eta'}{\sqrt{\beta_x}} \\
0 & 0 & \nicefrac{1}{\sqrt{\beta_y}} & 0 & 0 & 0 \\
0 & 0 & \nicefrac{\alpha_x}{\sqrt{\beta_y}} & \sqrt{\beta_y} & 0 & 0 \\
\eta' & \eta & 0 & 0 & 1 & 0 \\
0 & 0 & 0 & 0 & 0 & 1
\end{array}
\right )
\end{equation}
where $\beta$ and $\alpha$ are the Twiss parameters, and $\eta$ is the dispersion function at the given point in the lattice. This neglects effects like linear transverse coupling, vertical dispersion, and RF oscillations. Because IOTA will only have blocking RF, this is unnecessary for the initial ring design. However, future work must consider the effects of accelerating RF and synchrotron oscillations in greater detail.

The map $\mathcal{A}$ is a constant at one particular location of the ring. We will need to manipulate $\mathcal{A}$ to move around the ring, particularly when we are to concatenate all of the elliptic potential steps on the edges of eqn.~\ref{iotaMap1}. So, how do we move from one point on the ring to the other?

The answer is to look at the product in eqn.~\ref{iotaProd}. If we want to know the transfer map from element $M$ to element $M-1$, instead of from element $0$ to element $M$, this is just a cyclic permutation of the product. To enact this cyclic permutation, we multiply the inside and outside by the transfer map to get from element $1$ to element $M$ and its inverse:
\begin{equation}
e^{-\lieop{H_0'} C} = e^{-\lieop{h_M} \ell_M} \left (\prod_{i = 0}^M e^{-\lieop{h_i} \ell_i}\right ) e^{\lieop{h_M} \ell_M}
\end{equation}
This then translates in the normal co\"{o}rdinate representation as
\begin{equation}
e^{-\lieop{H_0'}C} =  e^{-\lieop{h_M} \ell_M}  \mathcal{A}^{-1} e^{-\lieop{\overline{H}}} \mathcal{A}e^{\lieop{h_M} \ell_M}
\end{equation}
In the section discussing normalizing maps earlier, we noted that $\overline{H}$ needs to be a quantity invariant of the particular location in the ring. Thus, the normalizing map must be what changes under this change of azimuth. Hence, if $\mathcal{A}$ is the normalizing map at the $M^{th}$ element, then
\begin{equation}
\mathcal{A}_M = \mathcal{A}e^{\lieop{h_M} \ell_M}
\end{equation}
must be the normalizing map of the $(M-1)^{st}$ element.

We will assume a rather generic form for $\overline{H}$, since it could contain any number of complex elements. We have, however, restricted ourselves to not having any elements which change $\delta$, such as an RF cavity. Assuming we have incorporated the chromaticity correction calculations in Appendix A into our computation, we are left with
\begin{equation}
\begin{split}
\overline{H} =& \frac{\mu_0}{2} \bigl \{  [1 - C_x (\delta)] \left (\overline{p}_x^2 + \overline{x}^2 \right )  \\
&+ [1 - C_y(\delta)] \left (\overline{p}_y^2 + \overline{y}^2 \right ) \bigr \}  + \frac{1}{2} \alpha_C \delta^2 + \textrm{h.o.t.}
\end{split}
\end{equation}
where the higher order terms include sextupole, octupole, \emph{etc.} terms left over from the chromatic correction scheme. Thus, $C_x(\delta)$ and $C_y(\delta)$ are assumed to include these corrections. In the absence of such corrections, we are left with the bare lattice chromaticities. These higher order terms will affect the dynamic aperture and break the integrability of the system, as they do in the case of conventional linear strong focusing lattices. They are higher order, though, and we are focused here on the lowest order of the nonlinear integrable lattices first.

As one final note, we can rewrite $\mathcal{A}^{-1}$ in a more illustrative form for the cancellation of the beta functions in the elliptic potential. The upper-left $4\times4$ matrix contains transformations on co\"{o}rdinates for on-momentum dynamics, and therefore does not mix in the longitudinal momentum offset through dispersion. We thus write
\begin{equation}\label{normalizingMap}
\mathcal{A}^{-1} = \begin{pmatrix}
 \multicolumn{4}{c}{\multirow{4}{*}{$\mathbb{A}^{-1}$}}& 0 & - \nicefrac{\eta}{\sqrt{\beta_x}}\\
 \multicolumn{4}{c}{}& 0 & - \nicefrac{\alpha_x \eta + \beta_x \eta'}{\sqrt{\beta_x}} \\
 \multicolumn{4}{c}{}  & 0 & 0 \\
 \multicolumn{4}{c}{}  & 0 & 0 \\
\eta' & \eta & 0 & 0 & 1 & 0 \\
0 & 0 & 0 & 0 & 0 & 1
\end{pmatrix}
\end{equation}
where $\mathbb{A}^{-1}$ is the upper-left $4\times4$ matrix, containing the Twiss parameters $\alpha$ and $\beta$. Later, we will freely interchange $\mathbb{A}^{-1}$ as either the upper $4\times4$ matrix, or a $6\times 6$ with the lower right $2\times2$ as the identity. This notation will prove convenient in the next section, when we look at the careful cancellation of $\beta$ for generating the integrable potentials used by Danilov and Nagaitsev in terms of normalizing maps.

\section{The IOTA Transfer Map, and an Integrable Hamiltonian}

In the previous section, we discussed the transfer map approach to the bare IOTA lattice. This allowed us to create a normalizing co\"{o}rdinate transformation $\mathcal{A}$ and rewrite the IOTA transfer Hamiltonian in the quadratic form, plus chromaticity, plus remaining nonlinearities that arise due to the chromatic correction schemes. The goal here is to do for the outer products of the IOTA transfer map what was done for the bare lattice -- obtain an invariant quantity and the normalizing co\"{o}rdinates in which it is conserved. This will yield the single-turn Hamiltonian and associated invariants, along with describing the effects of dispersion on the integrability of the single-turn Hamiltonian (see Appendix B for more details).

We begin from the transfer map in eqn.~\ref{iotaMap1}, but with the new map for the bare IOTA ring plus the normalizing map:
\begin{widetext}
\begin{equation} \label{iotaMap1}
\begin{split}
\mathcal{M}_{\textrm{IOTA}} = \prod_{i = \nicefrac{N}{2}+1}^N \exp \left \{ - \lieop{ \frac{t_i}{1 + \delta} e^{-\lieop{\nicefrac{\vec{p}^2}{2}} (i - \nicefrac{N}{2} + \nicefrac{1}{2})\Delta s} U_i(x, y) \Delta s}  \right \} \times\\
\left (\mathcal{A}^{-1} e^{-\lieop{\overline{H}} C} \mathcal{A} \right)& \times\\
\prod_{i = 0}^{\nicefrac{N}{2}} \exp &\left \{ - \lieop{ \frac{t_i}{1 + \delta} e^{-\lieop{\nicefrac{\vec{p}^2}{2}} (i + \nicefrac{1}{2}) \Delta s} U_i(x, y) \Delta s} \right \}
\end{split}
\end{equation}
\end{widetext}
We will try to move the $\mathcal{A}$ and $\mathcal{A}^{-1}$ operators to the outside of $\mathcal{M}_{\textrm{IOTA}}$ in the same way that we moved the drifts into the middle of the map. As before, we will look at the $\prod_{i=0}^{\nicefrac{N}{2}}$ half, as the argument is symmetrical.

We look the first term in the product:
\begin{equation}
\mathcal{A} \exp \left \{ - \lieop{\frac{t_0}{1 + \delta} e^{-\lieop{\nicefrac{\vec{p}^2}{2}} \nicefrac{ \Delta s}{2}} U_0(x, y)} \right \}
\end{equation}
Here we do two things: we insert an identity on the right side, and we take the drift component out of the exponent using the same identity we used to put it there in the first place. This process gives us:
\begin{equation}
\begin{split}
\mathcal{A} \exp \left \{ - \lieop{\frac{t_0}{1 + \delta} e^{-\lieop{\nicefrac{\vec{p}^2}{2}} \nicefrac{\Delta s}{2}} U_0(x, y)} \right \} = \\
\mathcal{A}_0\exp \left \{ - \lieop{\frac{t_0}{1 + \delta} U_0(x, y) \nicefrac{\Delta s}{2}} \right \} 
\mathcal{A}_0^{-1} \mathcal{A}
\end{split}
\end{equation}
where
\begin{equation}
\mathcal{A}_0 = \mathcal{A} e^{-\lieop{\nicefrac{\vec{p}^2}{2}} \nicefrac{\Delta s}{2}}
\end{equation}
is the normalizing map after the $0^{th}$ half-drift. In general, from our previous discussion of how the normalizing map transforms, we can write 
\begin{equation}
\mathcal{A}_i = \mathcal{A} e^{-\lieop{\nicefrac{\vec{p}^2}{2}} (i +\nicefrac{1}{2}) \Delta s}
\end{equation}
so long as we are inside the drift region. This is familiar from the usual way that the Twiss parameters are said to transform in linear lattices~(\cite{sylee_text:04}, Chap. 2, \S II.2). By repeating this operation, we are able to slip $\mathcal{A}$ all the way to the right. This leaves us with the product:
\begin{equation}
\begin{split}
\mathcal{A} &\left [ \prod_{i = 0}^{\nicefrac{N}{2}} \exp \left \{ - \lieop{ \frac{t_i}{1 + \delta} e^{-\lieop{\nicefrac{\vec{p}^2}{2}} (i + \nicefrac{1}{2}) \Delta s} U_i(x, y) } \Delta s\right \} \right ]=\\
&\left [\prod_{i = 0}^{\nicefrac{N}{2}} \mathcal{A}_i \exp \left \{ - \lieop{ \frac{t_i}{1 + \delta}  U_i(x, y) }\Delta s  \right \}\mathcal{A}_i^{-1} \right ]\mathcal{A}
\end{split}
\end{equation}

Now we note that the products are of a Lie map which normalizes the co\"{o}rdinates and a thin transfer map for the elliptic element. Using the similarity transformation rule in a way we have grown accustomed to, we write this as:
\begin{equation}
\begin{split}
\mathcal{A}_i \exp \left \{ - \lieop{ \frac{t_i}{1 + \delta}  U_i(x, y) }\Delta s  \right \}\mathcal{A}_i^{-1} =\\
\exp \left \{ - \lieop{ \frac{t_i}{1 + \delta}  \mathcal{A} \circ U_i(x, y) }\Delta s  \right \}
\end{split}
\end{equation}
Now the reasoning for rewriting the normalizing map in the suggestive way we did in eqn.~\ref{normalizingMap} should begin to be apparent.

We have thus packed up, unpacked, and then packed up again the IOTA transfer map to obtain its normalized co\"{o}rdinates including longitudinal momentum spread. The condition of matched beta functions and their cancellation in the original work by Danilov and Nagaitsev~\cite{danilovNagaitsev:2010} is the same as saying that
\begin{equation}
t_i \mathbb{A}_i \circ U_i(x, y) = \frac{t}{\beta_i}~ \mathcal{U}(\overline{x}, \overline{y})
\end{equation}
\emph{independent of the position}. That is to say, for the case when $\delta=0$, $t~\mathcal{U}(\overline{x}, \overline{y}) = \beta_i t_i U_i(\mathbb{A}^{-1} z)$ The addition of the dispersion for off-momentum particles modifies this so that
\begin{equation}
t_i \mathcal{A}_i \circ U_i(x, y) = \frac{t}{\beta_i}~ \mathcal{U}\left ( \overline{x} -  \eta_i \delta, \overline{y} \right )
\end{equation}
where $\eta_i$ is the value of the dispersion function at the $i^{th}$ slice and $\delta$ is the relative energy deviation. Because all of these operators commute with each other, we can collapse everything immediately to get the nonlinear integrable transfer map:
\begin{widetext}
\begin{equation}
\mathcal{M}_{\textrm{IOTA}} = 
\mathcal{A}^{-1} \exp \left \{-\lieop{\frac{t}{1 + \delta} \int_{\nicefrac{\ell}{2}}^\ell ds ~\frac{1}{\beta(s)} \mathcal{U}\left (\overline{x} - \frac{\eta(s)  \delta}{\sqrt{\beta(s)}}, \overline{y} \right ) } \right \}
 e^{-\lieop{\overline{H}}} 
 \exp \left \{-\lieop{\frac{t}{1 + \delta} \int_{0}^{\nicefrac{\ell}{2}} ds ~\frac{1}{\beta(s)}\mathcal{U}\left (\overline{x} - \frac{\eta(s)  \delta}{\sqrt{\beta(s)}}, \overline{y} \right )} \right \}   \mathcal{A}
\end{equation}
\end{widetext}
This yields the total single-turn Hamiltonian
\begin{widetext}
\begin{equation}
\begin{split}
\overline{\mathcal{H}} = \frac{\mu_0}{2} \left \{  [1 - C_x (\delta)] \left (\overline{p}_x^2 +  \overline{x}^2 \right ) + [1 - C_y (\delta)] \left (\overline{p}_y^2 + \overline{y}^2 \right )    \right \}+ 
\frac{t}{1 + \delta} \int_{0}^{\ell_{\textrm{drift}}} \frac{1}{\beta(s')}\mathcal{U} \left (\overline{x} - \frac{\eta(s')  \delta}{\sqrt{\beta(s')}} , \overline{y} \right) ds' + \frac{1}{2} \alpha_C \delta^2+ \textrm{h.o.t.}
\end{split}
\end{equation}
\end{widetext}
where the higher order terms include remnants of the chromatic correction scheme, Poisson brackets between the elliptic potential and the chromatic correction higher order terms, \emph{etc.} These terms are important in understanding things like the dynamic aperture at higher order, and the loss of integrability due to the chromatic correction schemes. However, we must get the zeroth-order Hamiltonian correct before worrying about how perturbations affect it.

Looking at this Hamiltonian, and comparing to the integrable Hamiltonian in two dimensions discussed in Appendix \ref{bdEqn}, we note two things:
\begin{enumerate}
\item Chromaticity differences break the assumptions in the derivation of the Bertrand-Darboux equation
\item Dispersion breaks the form of the elliptic potential for off-momentum particles
\end{enumerate}
The latter of the two is straightforward enough to correct. Dispersion-free sections are a standard tool in lattices, and indeed the current design for the IOTA ring has dispersion-free sections in the drifts where the nonlinear elliptic magnets will be placed.

The former is more interesting. In a conventional linear strong focusing storage ring, chromaticity and its correction is a balancing act between keeping the tune spread sufficiently small to avoid crossing low order resonance lines, while keeping it large enough to maintain enough Landau damping to prevent coherent instabilities. In the absence of any collective effects, avoiding nonlinear resonances requires making the chromaticity as small as possible.

The nonlinear integrable optics already has enormous tune spreads -- that is the reason it was designed in the first place. The tune spreads already cross low order resonance lines. Chromaticity correction should then be focused not on avoiding resonance lines, but on restoring integrability to off-momentum particles. Therefore, one would want to keep the higher order terms as small as possible, to avoid their impinging on the integrability. This means we must tune our sextupoles, octupoles, decapoles,..., to make the vertical and horizontal chromaticity equal.

If we define $C_0(\delta) = (C_x(\delta) + C_y(\delta))/2$ and $\Delta_C = (C_x(\delta) - C_y(\delta))/2$, and we assume the linear dispersion vanishes, we have 
\begin{equation}
\overline{\mathcal{H}} = \overline{\mathcal{H}}_0 + \Delta_C \left ( \overline{p}_x^2 + \overline{x}^2 - \overline{p}_y^2 - \overline{y}^2 \right )
\end{equation}
where $\overline{\mathcal{H}}_0$ is the ``ideal" Hamiltonian described in~\cite{danilovNagaitsev:2010} to within factors including $C_0(\delta)$. This makes explicit that all that is required of our chromaticity correction scheme is enough nonlinearity to make $\Delta_C$ vanish.

By building a lattice which satisfies these two considerations, we are left with a single-turn Hamiltonian (neglecting higher order terms here)
\begin{equation}
\overline{\mathcal{H}} = \frac{\mu_0}{2} \left [1 - C(\delta) \right ] \left ( \overline{p}_x^2 + \overline{x}^2 + \overline{p}_y^2 + \overline{y}^2 \right ) + \nu_0 ~t~ \mathcal{U}(\overline{x}, \overline{y})
\end{equation}
where $\nu_0 = \int_0^\ell ds' \frac{1}{\beta(s')}$ is the phase advance across the drift where the elliptic elements are placed. This form is similar to the form for the Hamiltonian in~\cite{danilovNagaitsev:2010}. Note that
\begin{equation}
e^{-2 \pi\lieop{ (\nicefrac{\overline{p}_x^2}{2} + \nicefrac{\overline{x}^2}{2})}} \overline{z} = \overline{z}
\end{equation}
for the phase space co\"{o}rdinates $\overline{z}$, and similarly for $y$ and $p_y$. This is simply to say that this particular operation is a $2 \pi$ rotation in the normalized variables, which is just the identity. Current design efforts for the IOTA ring are such that $\mu_0 = 2 \pi N + \nu_0$, which is to say that there is a $2 \pi N$ phase advance between the end of the elliptic element and the beginning.

Using this design and factoring out the $2 \pi N$ rotation, the single turn transfer map takes the form
\begin{equation}
\mathcal{M} = \mathcal{A}^{-1} e^{- \frac{\nu_0}{2} \lieop{\overline{\mathcal{H}}}} \mathcal{A}
\end{equation}
with the normalized Hamiltonian generating the map given by
\begin{equation}
\overline{\mathcal{H}} = \frac{1}{2} \left (1 - \frac{ \mu_0 C(\delta)}{\nu_0} \right ) \left ( \overline{p}_x^2 + \overline{x}^2 + \overline{p}_y^2 + \overline{y}^2 \right ) + t~ \mathcal{U}(\overline{x}, \overline{y}) 
\end{equation}
which, for on-momentum particles, is identical to the invariant Hamiltonian derived in~\cite{danilovNagaitsev:2010}.

\section{Discussion \& Conclusions}

The original work of Danilov and Nagaitsev, presenting the elliptic potential as a way of having large tune spreads in integrable dynamics, represents a zeroth order design of a realizable nonlinear integrable lattice. This work, considering a coasting beam with energy spread, chromaticity, and dispersion, represents a first order correction, illustrating the first operational principles for a realistic beam. Many other key issues remain -- the effects of magnet errors, synchrotron oscillations, and nonlinear synchrobetatron coupling on the dynamic aperture to name a few -- but fundamental design principles for nonlinear integrable lattices are taking shape.

In this paper, we have reached three important conclusions about nonlinear integrable optics. The first is that conventional notions of chromaticity survive, and existing correction schemes are perfectly valid and useful. We have also shown that the elliptic elements should be in dispersion-free sections. Finally, we have concluded that chromaticity breaks the integrability of nonlinear integrable lattices, but that a chromatic correction scheme which tunes the chromaticities to be equal will restore integrable single-particle motion. Because it does not matter whether $C_x$ or $C_y$ are tweaked, this leaves the lattice designer with much freedom in designing the correction scheme.

We have considered these chromatic correction schemes to quadratic order in the transverse dynamics and arbitrary order in $\delta$, i.e. to order $\delta^n z^2$, while neglecting terms of order $z^3$ or higher in the Tier 1 and Tier 2 lattices, assuming that their co\"{e}fficients are small compared to the elliptic potential. These terms affect the dynamic aperture, as they do in chromaticity correction schemes in linear strong focusing lattices. However, we have not considered their effect on the dynamic aperture in this study.

The primary approximation of this work is the assumption of a coasting beam. When synchrotron oscillations are present, the chromaticity will create coupling with the highly nonlinear transverse dynamics. Because the synchrotron tune is typically so much smaller than the transverse tunes, one might expect that an adiabatically conserved quantity would remain even with the nonlinear coupling. How this affects the dynamic aperture remains to be seen and is necessary to understand before a fully functional integrable optics machine could be built.

\section{Acknowledgements}

The authors would like to thank \'{E}. Forest (KEK) for useful discussions. This material is based upon work supported by the U.S. Department of Energy, Office of Science, Office of High Energy Physics under Award Number DE-SC0011340.

\appendix

\section{Lie Operator Treatment of Sextupole Chromaticity Correction}

In the course of carrying out this research, we found that there is no complete write-up in the language of Lie operators on how sextupoles correct chromaticity in linear lattices. For completeness, we include a more detailed calculation here.

We start by considering a sample lattice given by the elements
\begin{equation}
\mathcal{M} = \mathcal{M}_{0\rightarrow1} \mathcal{M}_{\textrm{sext.1}} \mathcal{M}_{1\rightarrow2} \mathcal{M}_{\textrm{sext.2}}
\end{equation}
where $\mathcal{M}_{i i+1}$ is the transfer map for a purely linear lattice (drifts and quadrupoles only) from the $i^{th}$ to the $i+1^{st}$ location in the ring and $\mathcal{M}_{\textrm{sext.}}$ is a thin sextupole. The usual procedure for this thing (see Dragt~\cite{dragt_text:2011}, Forest~\cite{forest_text:1998}, or Chao~\cite{chao_lieAlgebraNotes:2009}) is to move all of the linear components to the left, then begin concatenating with the operators to the right using a perturbation series. This is done with the usual tricks we've carried out throughout this paper:
\begin{equation}
\begin{split}
\mathcal{M} = \mathcal{M}_{0\rightarrow1} &\mathcal{M}_{1\rightarrow2}  \times \\
& \exp \left \{ - \lieop {\mathcal{M}_{1\rightarrow2} V(x, y)} \right \} \exp \left \{ - \lieop { V(x, y)} \right \} 
 \end{split}
\end{equation}
where $V(x, y) = \frac{S_3}{3} (x^3 - 3 x y^2)$ is a thin sextupole potential. Now, if the composition above is a linear lattice, then it has the same normal form analysis we had before, with
\begin{equation}
\begin{split}
\mathcal{M} = \mathcal{A}^{-1} \exp \left \{ - \lieop{ \frac{\mu_x}{2} (1 - C_x \delta) J_x + \frac{\mu_y}{2} (1 - C_y \delta) J_y} \right \} \mathcal{A} \times \\
\underbrace{\exp \{-\lieop{\mathcal{M}_{1 \rightarrow 2}^{-1} V(x, y)}\}\exp \{-\lieop{ V(x, y)}\}}_{\exp \{-\lieop{V_{nl}} \}}
\end{split}
\end{equation}
where $J_x$ and $J_y$ are the usual action-angle variables. Let us now insert the identity in the form of $\mathcal{A}^{-1} \mathcal{A}$ between the two nonlinear operators, so now we have that
\begin{equation}
\begin{split}
\mathcal{M} = \mathcal{A}^{-1} \exp \left \{ - \lieop{ \frac{\mu_x}{2} (1 - C_x \delta) J_x + \frac{\mu_y}{2} (1 - C_y \delta) J_y} \right \} \times \\
\mathcal{A} \exp \{-\lieop{\mathcal{M}_{1 \rightarrow 2}^{-1} V(x, y)}\} \mathcal{A}^{-1} \mathcal{A}\exp \{-\lieop{ V(x, y)}\}\mathcal{A}^{-1} \mathcal{A}
\end{split}
\end{equation}
We can decompose $\mathcal{M}_{1 \rightarrow 2}^{-1}$ into a normal form, so long as the dynamics are integrable or, more specifically in this case, linear. Then we have that $\mathcal{M}_{1 \rightarrow 2} = \mathcal{B}_1^{-1} \mathcal{R} \mathcal{B}_2$, where $\mathcal{R}$ is a pure rotation, and $\mathcal{B}_i$ is the normalizing map at point $i$ in the lattice. Using the similarity transformation rules, this gives
\begin{equation}
\begin{split}
\mathcal{M} = \mathcal{A}^{-1} \biggl ( \exp \left \{ - \lieop{ \frac{\mu_x}{2} (1 - C_x \delta) J_x + \frac{\mu_y}{2} (1 - C_y \delta) J_y} \right \} \times \\
 \exp \{-\lieop{\mathcal{A}\mathcal{B}_2^{-1} \mathcal{R}^{-1} \mathcal{B}_1 V(x, y)}\} \exp \{-\lieop{ \mathcal{A} V(x, y)}\} \biggr )\mathcal{A}
\end{split}
\end{equation}
We note that $\mathcal{B}_2 = \mathcal{A}$ from our definitions, and hence those cancel. and we are left with the rotation and normalizing map. This leaves the final two Lie operators as
\begin{equation}
\begin{split}
 \exp \{-\lieop{\mathcal{A}\mathcal{B}_2^{-1} \mathcal{R}^{-1} \mathcal{B}_1 V(x, y)}\} \exp \{-\lieop{ \mathcal{A} V(x, y)}\} =\\ \exp \{ - \lieop{\mathcal{R}^{-1} V(\sqrt{\beta_x^1} \overline{x} - \delta \eta^1, \sqrt{\beta_y^1} \overline{y})} \} \times \\ \exp \{ - \lieop{ V(\sqrt{\beta_x^2} \overline{x} - \delta \eta^2, \sqrt{\beta_y^2} \overline{y})} \}
 \end{split}
\end{equation}
If the rotation $\mathcal{R}$ has its upper $4\times4$ as $\mathbb{R} = - \mathbb{I}$, i.e. a rotation of $(2k+1)\pi$ in both transverse directions, and the identity for the energy, this gives the final nonlinear potential as
\begin{equation}
V_{nl} = V \left (-\sqrt{\beta_x^1} \overline{x} - \delta \eta^1, -\sqrt{\beta_y^1} \overline{y} \right ) + V \left (\sqrt{\beta_x^2} \overline{x} - \delta \eta^2, \sqrt{\beta_y^2} \overline{y} \right)
\end{equation}
For sextupole elements with strengths $S_1$ and $S_2$, this gives, to first order in $\delta$,
\begin{equation}
\begin{split}
V_{nl} = \left ( -S_1 (\beta_x^1)^{\nicefrac{3}{2}} + S_2 (\beta_x^2)^{\nicefrac{3}{2}} \right ) \overline{x}^3 + \dots \\
3 \left ( S_1 \sqrt{\beta_x^1} \beta_y^1  - S_2 \sqrt{\beta_x^2} \beta_y^2 \right ) \overline{x} \overline{y}^2 + \\
3 \left ( S_1 \beta_x^1 \eta^1 + S_2 \beta_x^2 \eta^2 \right ) \overline{x}^2 \delta + 3 \left ( S_1 \beta_y^1 \eta^1 + S_2 \beta_y^2 \eta^2 \right ) \overline{y}^2 \delta
\end{split}
\end{equation}
We note that, generally, obtaining the desired phase advance may be difficult, so there may be a family of chromaticity-correcting nonlinear elements and rotations positioned to cancel the leading order in $\delta$ term. The first system of equations can be made to cancel exactly through an intelligent choice of $S_1$ and $S_2$. The second term tells us how much chromaticity can be cancelled. The concatenation can be done with the usual resonance basis method, described in \S2.4 of~\cite{forest_text:1998}. The completion of this perturbation theory analysis is left as an exercise to the reader.

\section{Derivation of the Bertrand-Darboux Equation from a Transfer Map Approach}\label{bdEqn}

Because the assumptions of the Bertrand-Darboux equation are so important to the conclusions of this paper, we here include a derivation of the result. A treatment in terms of constant Hamiltonians in time can be found in Whittaker~\cite{whittaker_text:1937}, \S 152. Since the theme of this paper is the use of Lie operators, we obtain the same result with this formalism instead.

We note that an invariant $I(p, q)$ of a single-turn map $e^{-\lieop{h}}$ is such that
\begin{equation}
e^{-\lieop{h}} I = I
\end{equation}
It is certainly sufficient that $\lieop{h} I = 0$ to satisfy this condition -- that is, $I$ commutes with the Hamiltonian Lie operator. We assume $h$ is quadratic in the momenta with no difference in coefficients between $p_x$ and $p_y$:
\begin{equation}
h = \frac{1}{2} \left (p_x^2 + p_y^2 \right ) + V(x, y)
\end{equation}
and that $I$ is a quadratic form of the momenta:
\begin{equation}
I(p, q) = P p_x^2 + Q p_x p_y + R p_y^2 + S p_y + T p_x + K
\end{equation}
where the coefficients may, in general, be functions of $x$ and $y$. Taking the Lie operator on $I$ and noting the relationship with Poisson brackets gives a set of ten partial differential equations corresponding to the ten coefficients of the third-order polynomial in $p_x$ and $p_y$. These coefficients must all vanish for the Poisson bracket to vanish generally.

We start with the first order coefficients, specifically
\begin{subequations}
\begin{equation}
2 P \frac{\partial V}{\partial x} + Q \frac{\partial V}{\partial y} - \frac{\partial K}{\partial x} = 0
\end{equation}
\begin{equation}
2 R \frac{\partial V}{\partial y} + Q \frac{\partial V}{\partial x} - \frac{\partial K}{\partial y} = 0
\end{equation}
\end{subequations}
We differentiate the first with respect to $y$ and the second with respect to $x$ and subtract to obtain the single differential equation
\begin{widetext}
\begin{equation}\label{bdBegin}
Q \left ( \frac{\partial^2 V}{\partial y^2} - \frac{\partial^2 V}{\partial x^2} \right ) + 2 (P - R) \frac{\partial^2 V}{\partial x \partial y} +\frac{\partial Q}{\partial y} \frac{\partial V}{\partial y} - \frac{\partial Q}{\partial x} \frac{\partial V}{\partial x} + 2 \left ( \frac{\partial P}{\partial y} \frac{\partial V}{\partial x} - \frac{\partial R}{\partial x} \frac{\partial V}{\partial y} \right ) = 0
\end{equation}
\end{widetext}
From the $p_x^3, p_y^3$ coefficients, we know that $P$ and $R$ are pure functions of $y$ and $x$, respectively. From the mixed cubic coefficients, $p_x^2 p_y$ and $p_y^2 p_x$ we have the relations
\begin{equation}
\frac{\partial Q}{\partial x} = - \frac{\partial P}{\partial y}
\end{equation}
and
\begin{equation}
\frac{\partial Q}{\partial y} = - \frac{\partial R}{\partial x}
\end{equation}
from which we can conclude that $Q = x y$, $P = - \nicefrac{1}{2}~ y^2 + c$ and $R = - \nicefrac{1}{2} ~x^2 + c'$. Substituting this all into the main equation yields the familiar Bertrand-Darboux result:
\begin{widetext}
\begin{equation}\label{bertrandDarbouxEqn}
x y \left ( \frac{\partial^2 V}{\partial x^2} - \frac{\partial^2 V}{\partial y^2} \right ) + \left ( y^2 - x^2 + C^2 \right ) \frac{\partial^2 V}{\partial x \partial y} + 3 y \frac{\partial V}{\partial x} - 3 x \frac{\partial V}{\partial y} = 0
\end{equation}
\end{widetext}
The corresponding invariant is obtained by inserting these results and continuing to back-solve the remaining five partial differential equations.

We note here that an important part of this derivation is that there are identical coefficients of $p_x^2$ and $p_y^2$ in the Hamiltonian. If that were not the case, then the coefficients of eqn.~\ref{bdBegin} would not be equal and the anisotropy would make the corresponding generalization of Bertrand-Darboux equation an explicit function of $\delta$. Thus, anisotropic chromatic effects would enter the potential and invariant in a way that cannot be canceled exactly. On the other hand, if the chromaticities are isotropic, then they can simply be factored out of the Hamiltonian thus:
\begin{equation}
h = \left ( 1 - \frac{\mu_0 C(\delta)}{\nu_0} \right ) \left [ \frac{1}{2} \left ( p_x^2 + p_y^2 +x^2 + y^2\right ) + \frac{1}{1 - \frac{\mu_0 C(\delta)}{\nu_0}} V(x, y) \right ]
\end{equation}
which is just a reparameterization of $V$ so long as $\delta$ is a constant. Thus, for a coasting beam in a ring with isotropic chromaticity, the invariant is restored by simply mapping $t \mapsto \nicefrac{t}{1 - \frac{\mu_0 C(\delta)}{\nu_0}}$.

\bibliography{iota_chromaticity.bib}

\end{document}